\documentclass[prb,aps,twocolumn,tightenlines,%
showpacs,floatfix]{revtex4}
\usepackage{graphicx}
\usepackage{amssymb}
\usepackage{amsmath}
\usepackage{colordvi}

\newcommand{\bgreek}[1]{\mbox{\boldmath$#1$\unboldmath}}

\begin{document}

\title{Spin relaxation in $n$-type ZnO quantum wells}

\author{C. L\"u}
\affiliation{Hefei National Laboratory for Physical Sciences at
  Microscale and Department of Physics, University of Science and Technology of China, Hefei,
  Anhui, 230026, China}
\author{J. L. Cheng}
\thanks{Author to whom correspondence should be addressed}
\email{jlcheng@mail.ustc.edu.cn.}
\affiliation{Hefei National Laboratory for Physical Sciences at
  Microscale and Department of Physics, University of Science and Technology of China, Hefei,
  Anhui, 230026, China}

\date{\today}
\begin{abstract}
We perform an investigation on the spin relaxation for $n$-type
ZnO (0001) quantum wells by numerically solving the
kinetic spin Bloch equations 
with all the relevant scattering explicitly included. We show the
temperature and electron density dependence of the spin relaxation
time under various conditions such as impurity density, well
width, and external electric field. We find a peak in the temperature 
dependence of the spin relaxation time at low impurity
density. This peak can survive even at 100~K, much higher than the
prediction and measurement value in GaAs. There also exhibits a
peak in the electron density dependence  
at low temperature. These two peaks originate from the nonmonotonic
temperature and  electron density dependence of 
the Coulomb scattering. The spin relaxation time can reach the order
of nanosecond at low temperature and high impurity density.
\end{abstract}

\maketitle

\section{Introduction}
Much attention has been devoted to the spin degree of the freedom of
carriers in the zinc oxide (ZnO) with wurtzite structure in
the last few years,\cite{ZnO} partly because of the very long spin
relaxation time (SRT)\cite{Ghosh,Gamelin,Lagrade} and the prediction
that ZnO can become ferromagnetic with a Curie temperature above room
temperature if doped with manganese.\cite{Ferrand} However, only few
works investigate on the spin dynamics properties of
ZnO: Experimentally, Ghosh \emph{et 
  al.}\cite{Ghosh} investigated the electron spin properties in
$n$-type bulk ZnO and discovered the electron spin
relaxation time varying from 20~ns to 190~ps when the
temperature increased from 10 to 280~K. Liu \emph{et
  al.}\cite{Gamelin} measured the SRT  in colloidal $n$-type ZnO
quantum dots could be as long
as 25~ns at room temperature by electron paramagnetic resonance
spectroscopy. Theoretically,
Harmon \emph{et al.} calculated the SRT in bulk material in the
framework of a single-particle model.\cite{Joynt} However, a
theoretical investigation on the SRT in quantum wells (QWs) is still
rare, which is our task in the present paper.

The spin relaxation can be induced by the following mechanisms:
(i) D'yakonov-Perel'(DP) spin relaxation mechanism,\cite{DPmech}
which is revealed to be the dominant mechanism in $n$-doped
semiconductors.\cite{Zhou,jjh} For $n$-doped ZnO QW, the spin-orbit coupling
(SOC) is at least one order of magnitude weaker compared with that of
the well studied semiconductor 
GaAs\cite{fu,Voon}, a simple estimation shows that the criterion of the strong
scattering\cite{Zhou} is always satisfied even the momentum scattering
is also weaker than that in GaAs due to the larger effective mass
$m^\ast$. Therefore the DP mechanism here can be described by the motional
narrowing picture\cite{Fabian} qualitatively and the induced SRT is 
\begin{equation}
\tau\propto \frac{1}{\langle
  \Omega_{\mathbf k}^2\rangle\tau_p}\ ,  
\label{eq:tau}
\end{equation}
with $\tau_p$ standing for the momentum
scattering time and $\langle
\Omega_{\mathbf k}^2\rangle$ for the inhomogeneous broadening induced
by the SOC. As  pointed out first by Wu {\it et al.}\cite{Wu1,Wu5,Weng} 
and then by Glazov and Ivchenko,\cite{Ivchenko}  the electron-electron
scattering has important contribution to the spin relaxation process.
Therefore $\tau_p$ used in Eq.\ (\ref{eq:tau}) should be revised as 
$\tau_p^{\ast}=\left[\left(\tau_p\right)^{-1}+\left(\tau_p^{ee}\right)^{-1}\right]^{-1}$
to include the electron-electron momentum relaxation time
$\tau_p^{ee}$. \cite{Wu1,Wu5,Weng,Zhou,jjh,Ivchenko,Glazov,Harley}
(ii) Elliott-Yafet 
mechanism.\cite{EYmech} The revised 
criterion in Eq. (22) of Refs. [\onlinecite{jjh}] gives 
$\Theta\approx 340$~eV, which means the Elliott-Yafet mechanism is
negligible compared with the DP mechanism
\cite{Joynt} due to the small spin split off energy, the large band gap,
and the large $m^{\ast}$. (iii) Bir-Aronov-Pikus mechanism,
\cite{BAPmech} which is always unimportant in 
$n$-doped semiconductor. \cite{Zhou,jjh} Therefore, we only investigate the
SRT induced by the DP mechanism for $n$-type ZnO QWs in the following. 

In this paper, we quantitatively calculate the SRT for $n$-type
ZnO QWs by using the fully microscopic spin kinetic Bloch equation (KSBE)
approach,  which has been successfully used in investigating the spin
relaxation in QWs\cite{Wu5,Weng,schu,Zhou,Wu1} and
in bulk semiconductors.\cite{jjh,shen} 
With all the relevant scattering included, the influence of   
temperature, electron density, impurity density, well width and 
electric field on the SRT are studied detailedly. The temperature and density
dependence of the SRT is shown to be nomonotonic, and we find that the
SRT increases with the electric field monotonically.

This paper is organized as follows: In Sec.~II we describe our model and the
KSBEs.  Our numerical results are presented in Sec.~III. We conclude
in  Sec.~IV.

\section{Model and KSBEs}

We start our investigation from a $n$-doped ZnO QW of well width $a$
grown in (0001) direction, considered to be $z$ axis. Due to the 
confinement of QW, the momentum states along $z$-axis is quantized by
subband index $n$. With the momentum vector $\mathbf k=(k_x,k_y)$ and the spin index $\sigma$, the electron Hamiltonian can be written as
$
H_e=\sum\limits_{n\mathbf
    k\atop \sigma_1\sigma_2}\left\{{\cal E}_{n\sigma_1\sigma_2\mathbf
    k} - e\mathbf E\cdot\mathbf R \delta_{\sigma_1\sigma_2}\right\}a^{\dag}_{n\sigma_1\mathbf
    k}a_{n\sigma_2\mathbf k} + H_{I}
  $. Here ${\cal E}_{n\sigma_1\sigma_2\mathbf
    k}=\varepsilon_{n\mathbf
    k}\delta_{\sigma_1\sigma_2} +  [\mathbf
  h_R(\mathbf k) +(\mathbf h_D)_n(\mathbf
  k)]\cdot\bgreek{\sigma}_{\sigma_1\sigma_2}$ with $\varepsilon_{n\mathbf k}=\frac{\mathbf k^2}{2m^{\ast}} + \frac{\langle
  k_z^2\rangle_{n}}{m^{\ast}}$ is the energy spectrum, $\mathbf
R=(x,y)$ is the position, the effective magnetic field given by the
Rashba due to the intrinsic wurtzite structure
inversion asymmetry and Dresselhaus SOC can be written as:~\cite{fu} 
\begin{eqnarray}
  \mathbf h_R(\mathbf{k}) &=& \alpha_e (k_y, -k_x, 0)\ ,\nonumber\\
  (\mathbf h_D)_{n}(\mathbf{k}) &=& \gamma_e(b\langle k_z^2\rangle_{n}
  - k_{\|}^2) (k_y, -k_x, 0)\ ,
  \label{eq:soc}
\end{eqnarray}
with $\alpha_e$, $\gamma_e$ and $b$ standing for the SOC
coefficients. $ \langle k_z^2  \rangle_n = \frac{\hbar^2 \pi^2
  n^2}{a^2}$ is the subband energy in a hard-wall confinement potential. 
The scattering Hamiltonian $H_I$ includes all the scatterings, such as
electron-nomagnetic impurity scattering, electron-phonon scattering, and
electron-electron  scattering. 

We construct the KSBEs in the collinear statistics by using the
non-equilibrium Green function method as follows:\cite{Wu0,Wu1,Weng,Wu5,Haug} 
\begin{equation}
  \label{Bloch_eq}
  \partial_t\rho_{\mathbf k} - e\mathbf
  E\cdot\bgreek{\nabla}_{\mathbf k}\rho_{\mathbf k}=
  \partial_t\rho_{\mathbf k}|_{\rm {coh}} +
  \partial_t\rho_{\mathbf k}|_{\rm {scat}}. 
\end{equation}
The density matrix $\rho_{\mathbf k}$ for momentum $\mathbf k$ is a
matrix with matrix elements $[\rho_{\mathbf
  k}]_{n_1\sigma_1;n_2\sigma_2}$ which include all the coherence
between different subbands and different spins. 
The second terms on the left-hand side of the kinetic equations
describe the electric field $\mathbf{E}$ driven effect. $\partial_t\rho_{\mathbf k}|_{\rm {coh}}$ is
the coherent term. $\partial_t\rho_{\mathbf k}|_{\rm {scat}}$
denotes the scattering, including the electron-impurity, the
electron-phonon, as well as the   electron-electron scattering. The
expressions for these terms are given in Appendix~A.

Before we give our numerical results, the qualitative analysis of the
DP mechanism due to the electron-electron scattering can be made at
strong scattering limit.  The perturbation theory shows
the effective electron-electron momentum 
scattering time $\tau_p^{ee}$ in degenerate and nondegenerate limits satisfies\cite{Vignale}
\begin{equation}
  \frac{1}{\tau_p^{ee}} \propto
  \begin{cases}
    T^2N_e^{-1}, & \mbox{for } T\ll T_F\ ,\\
    T^{-1}N_e, &\mbox{for } T\gg T_F\ ,
  \end{cases}
  \label{eq:tau_ee}
\end{equation}
which has nonmonotonic temperature $T$ and electron density $N_e$
dependence as the electron gas undergoes the transition from the
degenerate case to the nondegenerate case at the Fermi temperature
$T_F$.
As the SOC in ZnO QWs mainly depends on $\mathbf k$ linearly, the
inhomogeneous broadening is given by
\begin{equation}
  \langle
\Omega_{\mathbf k}^2\rangle \propto
\begin{cases}
  N_e, & \mbox{for } T\ll T_F\ ,\\
  T, & \mbox{for } T\gg T_F\ .
\end{cases}
\end{equation}
Then by Eq.\ (\ref{eq:tau}), the electron-electron scattering
contributes to the SRT $\tau$ as \cite{Harley}
\begin{equation}
  \tau\propto
  \begin{cases}
    T^2N_e^{-2}, & \mbox{for } T\ll T_F\ ,\\
    T^{-2}N_e, & \mbox{for } T\gg T_F\ .
  \end{cases}
\label{eq:tau_s}
\end{equation}
The SRT is expected to reach a minimum in $T$ dependence
or a maximum in $N_e$ dependence, and the location of the
extreme points satisfie $T\approx T_F$. 

\section{Numerical Results}

We numerically solve the KSBEs for the spin density
matrix $\rho$, from which we obtain the time evolution of the spin
polarization along $z$-direction:~\cite{Weng} 
\begin{eqnarray}
  \label{sum_S}
  P_z(t) = \sum_{n,\bf k}\{
  [\rho_{\mathbf k}]_{n\uparrow;n\uparrow}(t) - [\rho_{\mathbf
    k}]_{n\downarrow;n\downarrow}(t)\}/N_e\ ,
\end{eqnarray}
where $N_e$ is the total electron density. 
The SRT $\tau$ is extracted from the exponential decay of the
  envelope of $P_z(t)$.
The initial condition at
$t=0$ is taken to be 
\begin{eqnarray}
  \label{rho0}
[\rho_{\mathbf k}]_{n_1\sigma_1;n_2\sigma_2} =
f_{T,\mu_{\sigma_1}}(\mathbf k)\delta_{n_1n_2}\delta_{\sigma_1\sigma_2}\ .
\end{eqnarray}
Here $f_{T,\mu}(\mathbf k)=[1+e^{(\varepsilon_{\mathbf k}-\mu)/(k_BT)}]^{-1}$ gives the Fermi-Dirac
distribution. The spin dependent chemical potential $\mu_{\sigma}$ is chosen to satisfy
$P(0)=2.5\%$. The electron density and the quantum well width
are taken as $N_e=4\times10^{11}$/cm$^2$ and $a=10$~nm respectively unless
otherwise specified. All used parameters are listed in Table
\ref{parameter}. In the calculation, only the lowest two subbands are taken into account.

\begin{table}[ht]
\caption{Material parameters used in the calculation (from
   Ref.~\onlinecite{parameter} unless otherwise specified).}
\begin{center}
\begin{tabular}{lllllllllll}
  \hline \hline
  $\kappa_{\infty}$ & $3.7$
  &\mbox{}&\mbox{}&\mbox{}&\mbox{}&\mbox{}&\mbox{}&\mbox{}& $\kappa_0$
  & $7.8$ \\
  $m_e/m_0$ & $0.25$
  &\mbox{}&\mbox{}&\mbox{}&\mbox{}&\mbox{}&\mbox{}&\mbox{}&
  $v_{1}$~(km/s) & 6.08\\
  $v_{3}$~(km/s) & 6.09
  &\mbox{}&\mbox{}&\mbox{}&\mbox{}&\mbox{}&\mbox{}&\mbox{}&  
  $v_{4}$~(km/s) & 2.73 \\
  $v_{6}$~(km/s) & 2.79 
  &\mbox{}&\mbox{}&\mbox{}&\mbox{}&\mbox{}&\mbox{}&\mbox{}& b
  & 3.91$^a$  \\
  $\gamma_{e}$~(eV\AA$^3$) & 0.33$^a$
  &\mbox{}&\mbox{}&\mbox{}&\mbox{}&\mbox{}&\mbox{}&\mbox{}& 
  $\alpha_{e}$~(meV\AA)   & 1.1$^b$ \\
  $e_{15}$~(V/m) & $-0.35\times 10^9$
  &\mbox{}&\mbox{}&\mbox{}&\mbox{}&\mbox{}&\mbox{}&\mbox{}&
  $e_{33}$~(V/m) & 
  $1.56\times 10^9$\\
  \hline \hline
\end{tabular}
\end{center}
\label{parameter}
\hspace {-3pc} $^a$ Ref.~[\onlinecite{fu}]; \hspace {3pc} $^b$
Ref.~[\onlinecite{Voon}].
\end{table}

\subsection{Temperature dependence}

\begin{figure}[htp]
  \centering
  \includegraphics[height=5.5cm,angle=0]{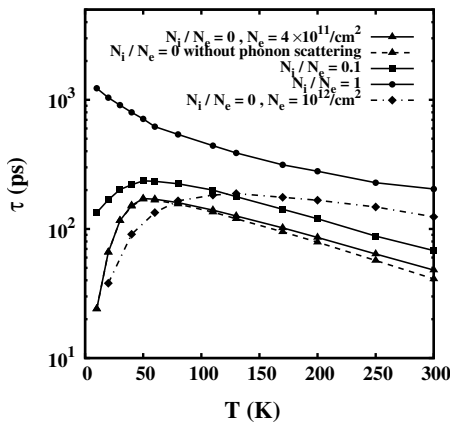}
  \caption{SRT $\tau$ {\em vs.} temperature $T$ at
different impurity densities. The dashed curve is obtained from the
calculation of excluding the electron-phonon scattering.}
  \label{fig:T}
\end{figure}

We now study the temperature dependence of the SRT presented in
Fig.~\ref{fig:T} for different impurity densities. 
The results are similar to that in GaAs QWs\cite{Zhou,Ji} and can be
understood as follows:  
(i) The SRT always increases with the impurity density $N_i$. 
It is because the system is in the strong scattering regime as
stated above due to the weak SOC, thus the spin
relaxation can be explained by the motional narrowing picture
qualitatively\cite{Fabian,Zhou}, and the additional scattering leads to longer SRT. 
(ii) The electron-phonon scattering is shown to be negligible over the whole
temperature regime by comparing the temperature
dependence of SRT with (solid curve  with 
$\blacktriangle$) or without (dashed curve  with 
$\blacktriangle$) the
electron-phonon scattering for the impurity free case in the same
figure.
(iii) The SRT presents a peak at very low $N_i$. In these cases, the
electron-electron scattering is the dominant 
scattering, therefore, as shown in Eq. (\ref{eq:tau_s}), the SRT shows a
maximum and the transition temperature $T_F$ for
$N_e=4\times10^{11}$~/cm$^2$ is about $44$~K, which is agree with our
numerical results. (iv) When the impurity
density is high enough such as $N_i = N_e$, the SRT decreases
monotonically with $T$. In this case the total scattering is mainly determined
by the impurity scattering, which depends weakly on the
temperature. However, the inhomogeneous broadening from the DP term
increases with the temperature, and results in shorter SRT. 

For GaAs QWs, the temperature peak of the SRT can only be
observed at low electron density ({\it i.e. low transition
  temperature}) and low impurity density\cite{Zhou,Ji}, because the
electron-phonon scattering becomes strong enough to destroy the nonmonotonic
$T$ dependence of the scattering time induced by the electron-electron
scattering. Such case can be avoid in ZnO QWs, in which the electron-phonon 
scattering is always pretty weak due to the large optical phonon energies
($\sim800$~K). Thus the temperature peak can be found even for high
electron density samples. One can easily find from the chained curve in
Fig.~\ref{fig:T}, which is calculated with parameters $N_e=10^{12}$~/cm$^2$ and
$N_i=0$, that the peak moves to $T\sim100K$. Therefore, the high
mobility ZnO QW is a good system for studying the electron-electron
scattering.

\subsection{Doping and well width dependence}

\begin{figure}[htp]
  \centering
  \includegraphics[height=5.5cm,angle=0]{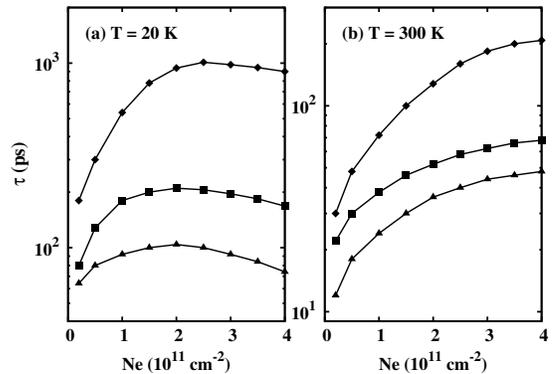}
  \caption{SRT {\em vs.} the electron density with different impurity
    densities and temperatures. (a) $T = 20$~K; (b) $T =
    300$~K. {$\blacktriangle$}: $N_i/N_e = 0$; {$\blacksquare$}:
      $N_i/N_e = 0.1$; {$\blacklozenge$}: $N_i/N_e = 1$.}
\end{figure}

Then we investigate the density dependence of the SRT
at different temperatures and impurity densities. In
Fig.~2 (a) we plot the SRT as a function of the electron
density with $T = 20$~K. One can see that for the low impurity density
case, the SRT reaches a maximum at $N_e \approx 2 \times
10^{11}$\ cm$^{-2}$, which has been 
pointed out in $n$-type bulk III-V semiconductors\cite{jjh,shen}. It
originates from the transition from the nondegenerate electron gas to the
degenerate electron gas and can be well explained by
Eq. (\ref{eq:tau_s}). Our calculation gives the transition density
of $2 \times 10^{11}$~cm$^{-2}$, corresponding
$T_F\sim22$~K, close to the lattice temperature of $20$~K. For the
case of $N_i = N_e$, the electron-impurity scattering time has the
same $N_e$ 
dependence as that for electron-electron scattering: in the
nondegenerate regime, $\frac{1}{\tau_p^{ei}}\propto N_i\langle U_q^2 
\rangle$, in which $\langle U_q^2 \rangle$ changes little;  in the
degenerate regime, $\frac{1}{\tau_p^{ei}}\sim N_i U_{k_f}^2 \propto
N_e/k_F^4 \propto N_e^{-1}$. Consequently, the peak still
exists and is almost at the same position. In comparison,
the SRT as a function of 
$N_e$ with $T = 300$~K is plotted in Fig.~2 (b). In this case,
one finds that the SRT increases
monotonically with $N_e$. This could be easily understood for that
$T_F \ll T$ is satisfied and it is in the nondegenerate regime, in
which the SRT increases with density as discussed above.

\begin{figure}
  \centering
  \includegraphics[height=5.5cm,angle=0]{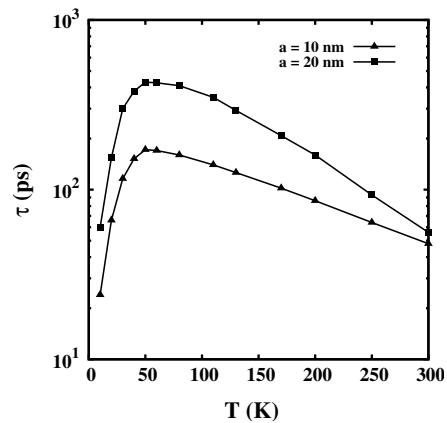}
  \caption{SRT {\em vs.} the temperature at different quantum
    well widths.}
\end{figure}

We further show the effect of quantum well width on the spin
relaxation. In Fig.~3 the SRTs versus temperature at
well widths $a = 10$~nm and $20$~nm are plotted
respectively. Both the SOC and the scattering\cite{Glazov,Harley} depend on the quantum
well width. However, comparing to the weak well width dependence of the
scattering, the fast decrease of $\langle k_z^2
\rangle$ in the DP term with $a$ dominates and so the SRT
increases with well width. 

\begin{figure}
  \centering
  \includegraphics[height=5.5cm,angle=0]{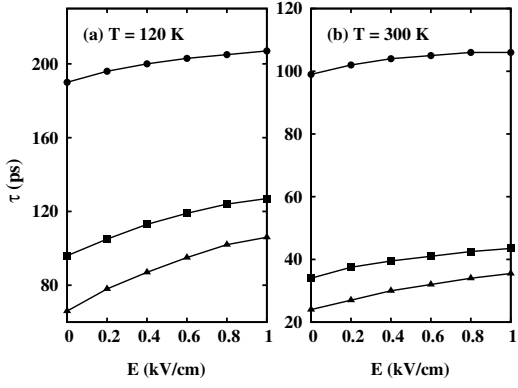}
  \caption{SRT {\em vs.} the electric field at different temperature
    and impurity densities. {$\blacktriangle$}: $N_i/N_e = 0$;
      {$\blacksquare$}: $N_i/N_e = 0.1$; {\large $\bullet$}: $N_i/N_e =
      1$.}
\end{figure}

\subsection{Electric field dependence}
Then we investigate the electric field dependence of the SRT
at different temperatures and impurity densities. 
In Fig.~4 we plot the SRT as a function of the 
electric field for different $T$. The electric
  field is applied along the $x$ axis. One can see that the SRT
increases monotonically for both low temperature and high temperature
cases. According to the previous investigation \cite{Wu5}, the
electric field will enhance both the momentum scattering due to the
hot-electron effect, and the inhomogeneous broadening due to the drift
of the electron distribution to larger $\mathbf k$ states. These two
effects are 
competing effects for the SRT: the former tends to enhance the
  SRT while the later
tends to suppress it. \cite{Wu5} For SOC with linear $\mathbf{k}$ 
dependence, the hot-electron effect dominates,\cite{jjh} thus the
SRT always increases with $E$.  

\section{Conclusion}

In conclusion, we have investigated the spin relaxation for $n$-type
ZnO (0001) QWs by numerically solving the KSBSs
with all the relevant scattering explicitly included. It is shown that
the electron-phonon scattering is pretty weak in ZnO
QWs, while the Coulomb scattering always plays an important
role. Therefore the ZnO QW is a good carrier to study the
electron-electron scattering. We find there exists a peak of
SRT both in the temperature dependence for a 
given electron density at low impurity density and in the electron
density dependence at low temperature. Both these two peaks originate
from the different temperature and electron density dependence
of $\tau^{ee}_p$ in degenerate and non-degenerate
case. Compared with the same effect in III-V semiconductor, 
\cite{Zhou,Ji,jjh,shen} this peak position can
occur at the temperature 
as high as 100~K and is easier to observe in experiments
due to the weak electron-phonon
scattering. When the impurity density
is high, the peak in the temperature dependence disappears and the SRT
decreases with temperature monotonously. Moreover, the peak
in the electron density dependence moves to larger electron
density which is beyond the scope of our interest when the temperature
is high.
We also investigate the hot-electron effect and show that the SRT always 
increases with the electric field. It is also
shown that the SRT reaches the order of nonosecond at low
temperature and high impurity density.

\begin{acknowledgments}
  The authors would like to thank M.W. Wu for proposing the topic as
  well as the directions during the investigation.
  This work was supported by the Natural Science Foundation of China
  under Grant No. 10725417, the National Basic Research Program of China
  under Grant No. 2006CB922005, and the Knowledge Innovation Project of
  the Chinese Academy of Sciences.  J.L.C was partially supported by
  China Postdoctoral Science Foundation.
\end{acknowledgments}

\begin{appendix}
\section{EXPRESSIONS FOR KINETIC BLOCH EQUATIONS}
Here we write the expressions for the coherent terms and the
scattering terms in the kinetic Bloch equations. The coherent terms in
Eq.~(\ref{Bloch_eq}) can be written as 
\begin{eqnarray}
  \label{coh}
  \partial_t\rho_{\mathbf k}\Big|_{\rm coh} = - i \Big[
   E_e(\bf{k}) + \sum_{\mathbf{Q}} V_{\mathbf{Q}}   I_{q_z}
   {\rho}_{\bf{k}-\bf{q}} I_{q_z} ,  \rho_{\bf{k}} \Big]\ , 
\end{eqnarray}
where $[A,B] = AB - BA$ denotes the commutator. $[E_e(\mathbf{k})]_{n_1\sigma_1;n_2\sigma_2} = {\cal
    E}_{n_1\sigma_1\sigma_2\mathbf k}\delta_{n_1n_2}$. The Coulomb Hartree-Fock term, which is always
negligible for small spin polarization, \cite{Weng, schu} is also included.
The form factor $I_{q_z} $ is also a matrix with matrix elements 
\begin{eqnarray}
  [I_q]_{n_1\sigma_1;n_2\sigma_2} =
 i a q  \delta_{\sigma_1,\sigma_2} [e^{i a q} \cos{\pi (n_1 -
    n_2)} -1 ] \nonumber \\ \times
  \left[\frac{1}{\pi^2(n_1 - n_2)^2 -
      a^2 q^2}
    - \frac{1}{\pi^2(n_1 + n_2)^2 - a^2 q^2}\right].
\end{eqnarray}
The statically screened Coulomb potential in the random-phase
approximation (RPA) reads~\cite{Haug} $V_{\bf Q} =
\frac{v_{Q}}{\epsilon(\mathbf q)}$ with the dielectric function
$\epsilon(\mathbf q)=1-
  \sum\limits_{q_z,\mathbf k\atop
    n_1;n_2}v_{Q}\big|[I_{q_z}]_{n_1;n_2}\big|^2
  \frac{f_{\sigma}(\varepsilon_{n_1,\mathbf{k+q}}) - f_{\sigma}(\varepsilon_{n_2,\mathbf k})}
{\varepsilon_{n_1,\bf k+q}
        - \varepsilon_{n_2,\bf k}}
$
and $\mathbf Q=(q_x,q_y,q_z)$. The bare Coulomb potential is $ v_Q =
4 \pi e^2 / Q^2 $. $f_{\sigma}(\varepsilon_{n,\mathbf{k}})=[\rho_{\mathbf
  k}]_{n\sigma;n\sigma}$.

The scattering terms in Eq.~(\ref{Bloch_eq}) can be
written as $\partial_t\rho_{{\bf{k}}}|_{\rm scat} =
\partial_t \rho_{{{\bf k}}}|_{\rm im} + 
\partial_t \rho_{{{\bf k}}}|_{\rm ph} +
\partial_t \rho_{{{\bf k}}}|_{\rm ee}$ in the Markovian limit with 
\begin{widetext}
  \begin{eqnarray}
    \partial_t\rho_{\mathbf k}\Big|_{\rm im} &=& \pi N_i \sum_{\mathbf{
        Q},n_1,n_2} |U^i_{\mathbf{Q}}|^2
    \delta(\varepsilon_{n_1,{{\bf k}-{\bf q}}} - \varepsilon_{n_2,{\bf k}})  I_{q_z} 
    [(1-{ \rho}_{{{\bf k}-{\bf q}}})T_{n_1}
    I_{-{q_z}} T_{n_2} {
      \rho}_{\bf k} -
    \rho_{{{\bf k}-{\bf q}}} T_{n_1} I_{-{q_z}} T_{n_2} (1-\rho_{\bf k})]  + h.c.\ , \nonumber  \\
    \partial_t\rho_{\mathbf k}\Big|_{\rm ph} &=&  \pi
    \sum_{{\mathbf{Q}},n_1,n_2,\lambda} 
    |M_{{\mathbf{Q}},\lambda}|^2 
    I_{q_z} \{
    \delta(\varepsilon_{n_1,{{\bf k}-{\bf q}}} - \varepsilon_{n_2,{\bf k}} +
    \omega_{{\mathbf{Q}},\lambda} ) [(N_{{\mathbf{Q}},\lambda} +1)(1-{
      \rho}_{{{\bf k}-{\bf q}}})  T_{n_1}I_{-{q_z}} T_{n_2} { \rho}_{\bf k} \nonumber \\
    &&\mbox{} - N_{\mathbf{
        Q},\lambda} \rho_{{{\bf k}-{\bf q}}}  T_{n_1} I_{-{q_z}} T_{n_2}(1-\rho_{\bf k})]
    +\delta(\varepsilon_{n_1,{{\bf k}-{\bf q}}} - \varepsilon_{n_2,{\bf k}} -
    \omega_{{\mathbf{Q}},\lambda} ) [N_{{\mathbf{Q}},\lambda}(1-{
      \rho}_{{{\bf k}-{\bf q}}})  T_{n_1}I_{-{q_z}}
    T_{n_2}{ \rho}_{\bf k}   \nonumber \\
  &&\mbox{}- (N_{\mathbf{
        Q},\lambda}+1) \rho_{{{\bf k}-{\bf q}}}  T_{n_1}I_{-{q_z}} T_{n_2}(1-\rho_{\bf k})]
    \} + h.c. \ ,  \nonumber \\
    \partial_t\rho_{\mathbf k}\Big|_{\rm ee} &=&  \pi \sum_{{\mathbf{q},q_zq_z^{\prime}},{\bf k}^{\prime}}
    \sum_{n_1,n_2,n_3,n_4} V_{\mathbf{Q}}V_{\mathbf Q^{\prime}}  \delta(\varepsilon_{n_1,{
        {\bf k}}-{\bf q}} - \varepsilon_{n_2,{\bf k}} + \varepsilon_{n_3,{\bf k}^{\prime}} -
    \varepsilon_{n_4, {\bf k}^{\prime} -{\bf q}})I_{q_z}
   \nonumber \\
  &&\mbox{}\times \{
    (1-{\rho}_{{{\bf k}-{\bf q}}})T_{n_1}
    I_{-{q_z}} T_{n_2}{   \rho}_{\bf k}
    \mbox{Tr}[(1-\rho_{{\bf k}^{\prime}})T_{n_3}
    I_{q_z^{\prime}} T_{n_4}\rho_{{\bf k}^{\prime} -{\bf q}}
    I_{-q_z^{\prime}}]  \nonumber \\
  &&\mbox{}- \rho_{\mathbf{
        k}-{\bf q}} T_{n_1} I_{-q_z} T_{n_2}(1-\rho_{\bf k})
    \mbox{Tr}[\rho_{{\bf k}^{\prime}}T_{n_3}
    I_{q_z^{\prime}}  T_{n_4} (1-\rho_{
        {\bf k}^{\prime}-{\bf q}})
    I_{-q_z^{\prime}} ]\}  + h.c. \ \ ,
  \label{scat}
\end{eqnarray}
\end{widetext}
in which $[T_{n_1}]_{n,n'} =  \delta_{n_1, n} \delta_{n_1,n'}$ and
$\mathbf Q^{\prime}=(q_x,q_y,q_z^{\prime})$. $N_i$ is the density of
impurities, and $ |U^{i}_{\mathbf{Q}}|^2 $ is the screened impurity
potential. $|M_{\mathbf{Q},\lambda}|^2$ and $N_{\mathbf{ Q},\lambda} =
[\mbox{exp}(\omega_{\mathbf{ Q},\lambda} /k_B T) - 1]^{-1}$ are the
matrix elements of the electron-phonon interaction and the Bose
distribution function respectively. $\omega_{\mathbf{ Q},\lambda}$
is the phonon energy spectrum. For the electron-phonon scattering, we
only include electron AC-phonon scattering, for which the explicit expressions can be found in
Refs.~\onlinecite{Lee} and \onlinecite{Rossi}. The electron-optical
phonon scattering is ignored for that the optical phonon energies
are around $ 70$ meV and are out of our temperature range.\cite{parameter}
\end{appendix}

\end{document}